# Effect of valence electrons on the core level x-ray photoelectron spectra of niobium oxide thin films prepared by molecular beam epitaxy

Jasnamol Palakkal, *,[1,2] Pia Henning,[1] and Lambert Alff[2]

[1]Institute of Materials Physics, Georg-August-University of Göttingen, Germany

[2]Institute of Materials Science, Technical University of Darmstadt, Germany

*E-mail: jpalakkal@uni-goettingen.de.

**Abstract**: X-ray photoelectron spectroscopy (XPS) is a versatile tool to identify an element's chemical and electronic state. However, the presence of various initial- and final-state effects makes the interpretation of XPS spectra tricky and erroneous. Metal oxides show multipeak XPS spectra when the charge carriers are coupled with the core hole of the photoion after photoemission. The extra peaks are often misconstrued to originate from other oxidation states of the cation. With a systematic approach to the analysis of Nb 3d XPS spectra of partially oxidized $NbO_x$ thin films prepared by molecular beam epitaxy, we found that the valence band electrons close to the Fermi level contribute to a satellite peak in Nb-oxides. Insulating Nb-oxide containing $Nb^{5+}$ ($d^0$) with no or fewer electrons near the Fermi level has no such peaks; on the other hand, $Nb^{5+}$ containing thin films exhibit this satellite peak when sufficient surface charge carriers contribute to the core hole-electron coupling.

**Graphical abstarct**

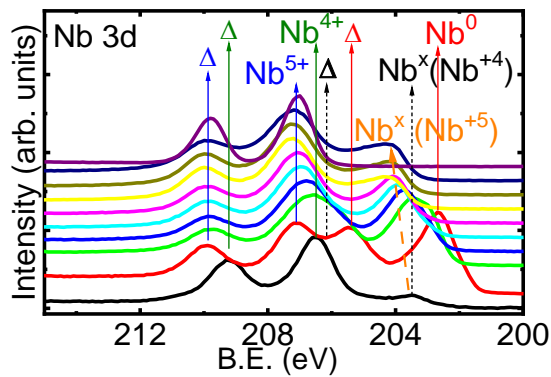

# 1. Introduction

The materials chemistry and physics community widely investigate Nb-based compounds because of their use in applied superconductivity, energy conversion devices, wireless communication systems, etc. [1-3] Identifying the exact electronic state of a cation is crucial when explaining and correlating the functionalities of a compound with its chemical state. A critical area where surface chemistry and electronic configuration are highly important is with respect to the catalytic activity of these materials. [1] X-ray photoelectron spectroscopy (XPS) has been used as an essential method to analyze the chemical and electronic state of a material's surface since the development of the technique. Many recent publications discuss XPS as an essential technique for identifying the oxidation state of Nb in $NbO_x$-based catalysts for their use in energy conversion and storage cells. [4-7]

XPS utilizes X-rays to induce the photoemission of electrons, which leaves a core hole behind. The spectral lines corresponding to the effects of the initial state of the element before photoionization are known as initial-state effects, and those effects induced by the perturbations resulting from photoemission are called final-state effects. [8] The ground-state polarization and spin-orbit splitting are associated with initial-state effects, while photoelectron-induced polarization and rearrangement effects are final-state effects. [8] The final-state effects cause spectral modifications or even the occurrence of additional spectral lines. This makes the interpretation of XPS spectra tiresome. Many XPS fitting software are available and being used by the scientific community. However, the software only provides a mathematical deconvolution of curves. The user should provide enough constraints to the curve fitting software to come up with a sensible analysis of the chemical state of the elements. This is why the community exclaims that fitting XPS spectra can be a 'tricky' job. A lack of knowledge and literature about the origin of different initial and final-state effects in XPS practice can create misinterpretations and wrong conclusions about the chemical state.

Lin *et al.* addressed this aspect in their work on transition metal (TM) oxides. [9] They concluded that the multiple peaks in the XPS spectra of charge-doped TM oxides do not always originate from different oxidation states but from the final-state effects. [9] In their work, they found that electron and hole doping create an additional peak at lower binding energy and higher binding energy, respectively, than the peak corresponding to the initial state of the cation. [9] It is likely that these additional peaks are often contemplated as originating from other oxidation states of the *TM* cation. [9, 10] When considering the final-state effects originating from particles/quasiparticles carrying electric charge, it is also important to see the surface electrical conductivity or the number of charge carriers of a material before analyzing its XPS spectra. Many defects during thin film deposition can introduce electron and hole doping in thin film material systems. Defect-induced conductivity can be present but comparatively less in powder samples. Because of this, the additional spectral features of a cation originating from final-state effects can be different when measured in powder and thin film samples.

Reports suggest that the XPS spectra of Nb 3d measured on various Nb-oxides exhibit multipeak structure. In 1977, Fontaine *et al.* measured the binding energies of $3d_{5/2}$ spectral line of $NbO_x$ as 202.4 eV for Nb ($Nb^0$), 204.5 for NbO ($Nb^{+2}$), 205.9 for $NbO_2$ ($Nb^{+4}$) and 207.4 for $Nb_2O_5$ ($Nb^{+5}$). [11] They suggested that the binding energy follows a linear relationship with the oxidation state and any binding energy between these values corresponds to a non-integer oxidation state. [11] Nb 3d core level XPS spectra of epitaxially grown single-phase $NbO_2$ films were reported to show an extra peak at a lower binding energy than that of $Nb^{4+}$,[12] which was corroborated as originating from an oxidation state of Nb between +2 and +4 by citing the work of Fontaine *et al.*. [11] Nb 3d XPS of $NbO_x$ thin films grown by molecular beam epitaxy (MBE) showed an asymmetric peak at lower binding energy along with $Nb^{5+}$ peak, which was assigned as originating from $Nb^{2+}$.[13] The lowest binding energy peak, assigned as $Nb^{2+,}$ also exhibits an asymmetry. [13] They used an Nb metal source and molecular oxygen to grow the Nb-oxide by MBE, but they didn't reflect on the presence of any unoxidized Nb in the samples. [13] NbO has a narrow phase formation region; that's why, generally, special synthesis methods are required to produce it ($Nb^{2+}$). [14, 15] NbO is reported to exist in a two-phase form typically, either together with Nb or with $NbO_2$. [14] In 1987, Halbritter reported the oxidation of Nb and found that Nb oxidation was dominated by the formation of $Nb_2O_5$, where Nb is a superconductor and $Nb_2O_5$ is an insulator. [16] Interestingly, the XPS spectra of Nb oxidized for 0.5 h showed peaks corresponding to five compounds, (i) Nb, (ii) $NbO_{0.2}$, (iii) NbO, (iv) $NbO_2$ and (v) $Nb_2O_5$.[16] Certainly it is challenging to look into the literature for confirming the oxidation state of Nb in such a scenario. Though the reports on the XPS spectra of Nb-based oxides are baffling, it is explicit that a systematic investigation of them is necessary to interpret the chemical and electronic state of these industrially important classes of materials.

Recently, we worked with SrNbO$_3$, a transparent conducting oxide (TCO), with a room temperature resistivity ($\rho$) of ~80 µΩcm and found that the XPS spectra of Nb 3d are showing an additional peak (labeled as Nb$^x$ in **Figure** 1 (a)) along with the peak of Nb$^{4+}$, which could be arising due to a final-state effect. Thapa *et al.* reported the Nb 3d spectra of SrNbO$_3$ capped with SrHfO$_3$. [17] They found that Nb$^{4+}$ and Nb$^{5+}$ are present in the uncapped samples, and with the thickness of the capping layer, the amount of Nb$^{4+}$ increased. [17] However, Nb$^{x+}$ was not shown in their results, which is likely due to the absence of sufficient charge carriers near the Fermi level of the uncapped SrNbO$_3$ sample as can be seen in the valence band XPS spectrum. [17] Though SrHfO$_3$ capped SrNbO$_3$ samples exhibit states near Fermi level, the surface capping layer and the overlapping Hf and Nb spectra make it challenging to recognize a final-state effect of Nb. [17] Sr$_{1-x}$NbO$_{3-\delta}$ powder samples with only Nb$^{4+}$ didn't exhibit this peak either. [18] We believe that a lack of surface charge carriers in the samples of these two works [17, 18] could be a reason for the absence of Nb$^{x+}$ peak, regarding it as originating from final-state effects.

Bigi *et al.* reported an overview of the electronic band structure of SrNbO$_3$, which has a $\rho$ of ~200 to 400 µΩcm, [19] with metallic conduction as in the case of the SrNbO$_3$ sample ($\rho$ of ~80 µΩcm) used in this work. They suggested heavier effective mass of charges and corresponding dynamical correlations in the sample. [19] Their work showed multipeak Nb 3d core-level spectra; however, the lower binding energy peak was assigned as originating from the Nb$^{2+}$ oxidation state impending from NbO, [19] as looked over before. Remarkably, their work showed a spectral contribution (S) at a higher binding energy than that of Nb$^{5+}$, which is sidelined to assign any particular valence states higher than Nb$^{5+}$ (4d$^0$). They reasoned that this S could be merged into Nb$^{5+}$ by adopting an asymmetric Doniach-Sunjic profile considering the metallicity of their sample. [19] However, the peak S is at a higher energy level than the core level; besides, we already know that the charge carriers are electrons in SrNbO$_3$, not holes. The electrons provide satellite features at a lower energy than the core level due to the core-hole-valence electron-induced attractive potential as reported by Lin *et al*. [9] We speculate that adding peaks corresponding to final-state Nb$^{x+}$ could have made the deconvolution of these spectra more straightforward.

The XPS spectra of SrMoO$_3$, another TCO, were reported to show additional spectra at a lower binding energy than that of Mo$^{4+}$, which was interpreted as a plasmon satellite, a final-state effect. [20] A recent theoretical calculation demonstrated that the plasmon satellites are spectrally narrow and relatively intense. [21] Both SrNbO$_3$ and SrMoO$_3$ are Pauli paramagnetic band metals. [22] The unpaired d$^1$ (or d$^2$) electrons of Nb$^{4+}$ (or Mo$^{4+}$) contributes to the conductivity of SrNbO$_3$ (or SrMoO$_3$). The possibility of a final-state effect due to plasmonic correlations, as in the case of SrMoO$_3$, is inevitable in SrNbO$_3$. However, we couldn't find any literature to support this argument since some works on SrNbO$_3$ didn't show this peak, and some other works corroborated this additional peak as originating from other oxidation states. It is worth mentioning that the TCOs SrNbO$_3$ and SrMoO$_3$ in thin film form show different metallic conductivity according to the defect concentration; in addition to that, the polycrystalline powder samples are less conducting or even non-conducting. [18-20, 22] We believe that the Nb$^{x+}$ peak in our SrNbO$_3$ samples could originate from final-state features due to the high crystalline quality and metallic conductivity of these films deposited by oxide MBE. To further prove this assumption and to locate the exact spectral position of the satellite peak originating due to the effect of the charge carriers of Nb$^{4+}$ containing materials, we systematically investigated the Nb 3d and valence band XPS spectra of a series of NbO$_x$ thin films. We compared them with the spectra of Nb-based perovskite oxide thin films of SrNbO$_3$ (metallic) and AgNbO$_3$ (insulator) prepared as part of other projects.

## 2. Experimental details

**Thin film preparation**. The partially oxidized NbO$_x$ thin films were grown on SrTiO$_3$ (111) substrates by oxide molecular beam epitaxy enabled by e-beam evaporation. The series of NbO$_x$ were prepared by varying the flow of oxygen from 0.00 to 0.35 SCCM (Samples: Nb_00 to Nb_35) that maintained a stable oxygen partial pressure in the growth chamber from 1x10$^{-11}$ to 1.7x10$^{-6}$ mbar, respectively. The deposition temperature was 500 $^{\circ}$C. The Nb pellets (dimension 1/4" Dia. x 1/2" Length) for e-beam evaporation were purchased from Kurt J. Lesker Company® (99.95% purity). The pellets were kept in a FABMATE® crucible insert with a copper crucible liner. The pellets were made entirely molten and solidified to form a single block of starter source before the start of each deposition. This provided a homogeneous evaporation rate throughout the deposition. The rate of Nb evaporation was controlled by varying the energy of the e-beam and monitored by using quartz crystal microbalance (QCM). Using a second QCM kept at the location of the substrate prior to deposition, we calibrated the QCM used for monitoring. A rate of 0.05 Å/s of Nb was maintained at the location of the substrate during the deposition. For all the samples, the deposition was stopped when the Nb metallic deposit at the monitoring QCM read 70 Å thickness of Nb, irrespective of the flow of oxygen.

**X-ray photoelectron spectroscopy (XPS)**. The XPS measurements were performed using a PHI Versaprobe 5000 spectrometer with Al $K_\alpha$ radiation. All the films were exposed to the atmosphere before performing the XPS measurements. Conducting screws and washers were used for clamping the films on the XPS sample holder to maintain a constant surface potential during XPS measurements. [23]

## 3. Results and discussions

We deposited a series of partially oxidized Nb-oxide ($NbO_x$) thin films by oxide molecular beam epitaxy (MBE). The samples were produced by partially oxidizing the elemental Nb during deposition by varying the flow of oxygen from 0.00 to 0.35 SCCM (Samples: Nb_00 to Nb_35). Other conditions were kept the same for all depositions (see experimental methods). We took the samples out of the growth chamber and exposed them to air before XPS measurements to introduce native oxide $Nb_2O_5$ on Nb metal. [16] The binding energies of Nb 3d spectra were corrected by taking C1s (=284.8 eV) as reference level. The O 1s spectra of all the $NbO_x$ samples (see **Figure S1**), after correcting the binding energy by taking C 1s as a reference, exhibit the same peak position, holding this calibration effective. O 1s spectra also show a shoulder peak between 532 and 533 eV, usually referred to as originating from surface adsorbed oxygen.[24-26] We noticed larger contributions to this peak in the sample prepared with the lowest flow of oxygen Nb_0.05. This peak gradually diminished upon increasing the oxygen flow, as expected when lowering the oxygen vacancy concentrations. The reported binding energies of Nb $3d_{5/2}$ spectral lines are listed in Table 1. [27, 28] We excluded the presence of NbN, NbC, and NbO in our films as we don't expect any nitride or carbonate formation during our MBE growth; further NbO is formed by the reduction of $Nb_2O_5$ by comproportionation, under special synthesis conditions. [15] (Avoiding the presence of NbO is a delicate topic as many reports already claimed the satellite peak as originating from NbO. We started the analysis of the spectra by considering the absence of NbO, which later made us arrive at a conclusion on the spectral positions of satellite peaks).

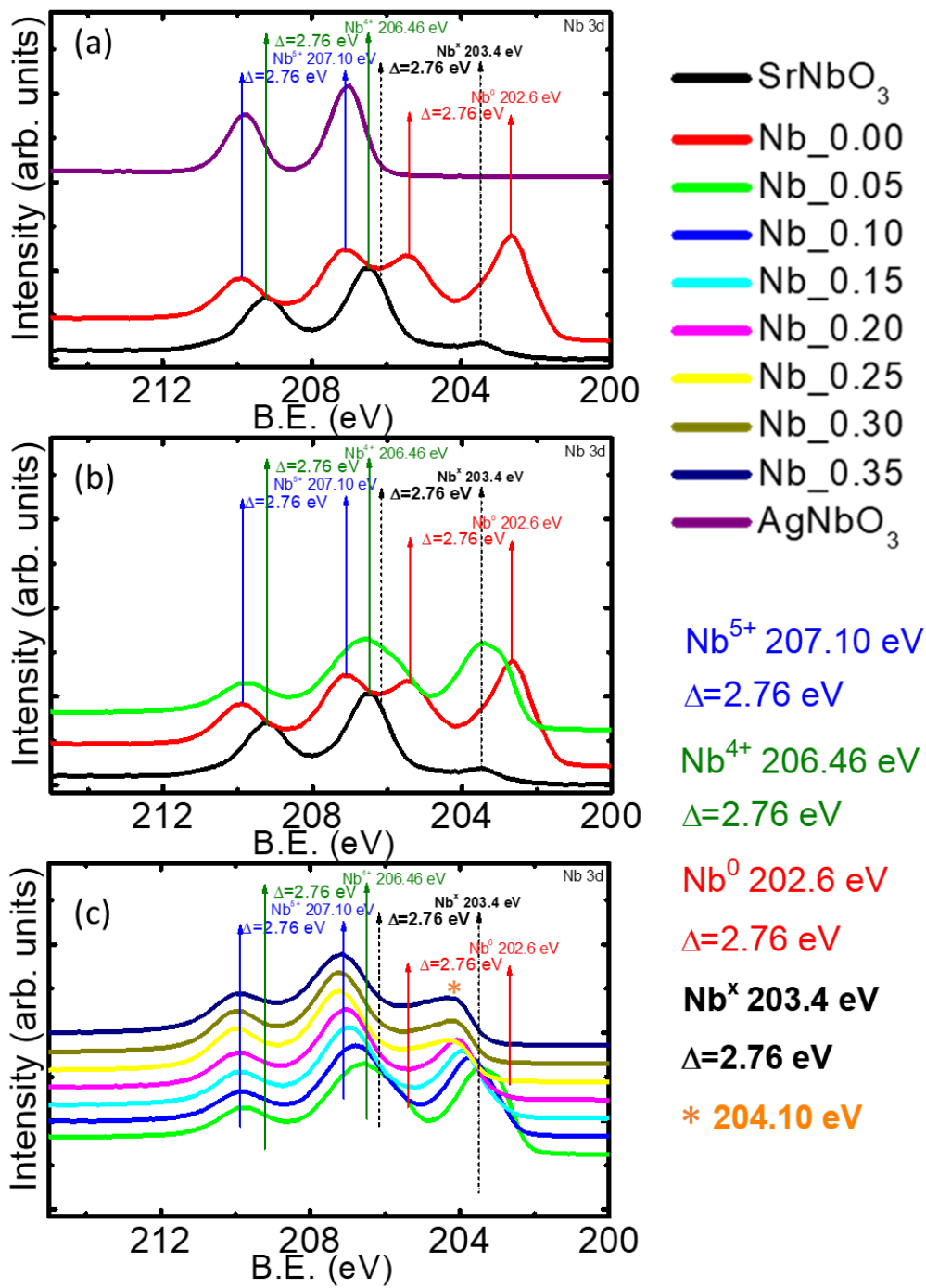

**Figure 1.** Nb 3d XPS spectra of various Nb-based samples (a) SrNbO$_3$, Nb_0.00, and AgNbO$_3$, (b) SrNbO$_3$, Nb_0.00, and Nb_0.05, and (c) Nb_0.05 to Nb_0.35.

**Table 1.** The binding energy (B.E.) values of Nb $3d_{5/2}$ spectral lines from NIST database [27] (as tabulated by the website [28])

| Sample | B.E. | Std. Dev. |
|---|---|---|
| Nb | 202.2 | 0.3 |
| $NbO_2$ | 206.2 | 1.0 |
| $Nb_2O_5$ | 207.4 | 0.4 |
| NbO | 203.7 | 1.0 |
| NbN | 203.6 | 0.2 |
| NbC | 203.7 | |

The Nb 3d XPS spectra of the series of $NbO_x$ samples from Nb_00 to Nb_0.35, together with $SrNbO_3$ and $AgNbO_3$, are plotted in **Figure 1** (a)-(c). The spectra of $SrNbO_3$, $AgNbO_3$, and Nb_00 are compared in **Figure** 1 (a). Nb_00 is a pure Nb metallic thin film grown with no oxygen background gas during deposition. Nb metal surfaces show the formation of native $Nb_2O_5$ when exposed to open air. [29, 30] All samples were exposed to air before XPS measurements to purposefully introduce this additional spectral line. Therefore, the native oxide formation (extrinsic $Nb^{5+}$) one-on-one corresponds to the amount of metallic Nb.

We observed the metallic $Nb^0$ $3d_{5/2}$ peak at ~202.6 eV with a spin-orbit splitting component $3d_{3/2}$ at a binding energy separation ($\Delta$) of ~2.76 eV. The secondary electronic excitations of Nb metal caused asymmetry in the core line of $Nb^0$ due to the core hole-valence electron coupling as per the Doniach Sunjic profile. [8] The $Nb^{5+}$ $3d_{5/2}$ peak of native oxide is at ~207.10 eV, confirmed by comparing the $Nb^{5+}$ $3d_{5/2}$ peak of $AgNbO_3$. The $Nb^{4+}$ $3d_{5/2}$ line of $SrNbO_3$ is at 206.46 eV. $Nb^{5+}$ and $Nb^{4+}$ have the same doublet separation ($\Delta$) of ~2.76 eV as $Nb^0$. The satellite peak $Nb^x$ of $SrNbO_3$ is at ~203.4 eV, which is clearly not part of the metallic $Nb^0$ peak.

Next, as shown in **Figure** 1 (b), we compared the Nb 3d spectra of $SrNbO_3$ and Nb_00 with that of Nb_0.05, the lowest oxidized $NbO_x$ sample in our series. Nb_0.05 has $Nb^{4+}$ spectral line, whereas Nb_00 didn't show any signatures of $Nb^{4+}$. Interestingly, Nb_0.05 shows the $Nb^x$ peak at ~203.4 eV as a broader shoulder at the low binding energy side, along with the presence of unoxidized $Nb^0$. Obviously, Nb_0.05 also has the extrinsic $Nb^{5+}$ since $Nb^0$ is still present in this sample. However, we are not discarding the possibilities of intrinsic $Nb^{5+}$ formation in Nb_0.05 during the growth since the oxygen flow was turned on, and some of the $Nb^{4+}$ formed could continue oxidizing to $Nb^{5+}$ during the deposition itself.

In **Figure** 1 (c), we compared the Nb 3d spectra of $NbO_x$ samples Nb_0.05 to Nb_0.35. From sample Nb_0.05 to Nb_0.35, we expect a gradual oxidation of $Nb^0 \rightarrow Nb^{4+}$ and $Nb^{4+} \rightarrow Nb^{5+}$ with a diminishing of extrinsic $Nb^{5+}$ corresponding to the decrease of metallic $Nb^0$. One can see that the $3d_{5/2}$ $Nb^x$ peak is shifting towards higher binding energy (towards the spectral line marked with an asterisk *) upon the increase in the flow of oxygen for samples Nb_0.05 to Nb_0.25. The peak position of $Nb^{4+}$ also shifts towards higher energy (towards the $Nb^{5+}$ line) for this set of samples due to the increase in the intrinsic $Nb^{5+}$ with the volume of flow of oxygen during the growth. The peaks marked with an asterisk * and the $Nb^{5+}$ peak are not shifting for samples Nb_0.25 to Nb_0.35. This confirms that upon oxidation, the $Nb^{4+}$ is converting to $Nb^{5+}$, and correspondingly, the peak at $Nb^x$ is moving towards the higher binding energy. In principle, one could deconvolute the $Nb^x$ peak to two components, corresponding to $Nb^{4+}$ and $Nb^{5+}$. Though assuming $Nb^x$ as originating due to the final-state effect of charge carriers from $4d^1$ is acceptable, we were perplexed by the appearance of such an additional satellite peak for $Nb^{5+}$ ($4d^0$), which has no unpaired electrons for its contribution to the final-state plasmonic effect. This satellite peak was absent for insulating $AgNbO_3$, where only $Nb^{5+}$ exists.

It became apparent when we checked the valence band XPS spectra of the Nb oxides, as shown in **Figure** 2, that there are charges in all the Nb oxides near the Fermi energy level except for $AgNbO_3$. The density of states (DOS) near the Fermi level for $SrNbO_3$ is small compared to other oxides due to only one d electron present, but it is the closest to the Fermi level, affirming the material's metallic nature. The peak of DOS of $NbO_x$ lies around 1.5 eV, proclaiming a metallic to semiconducting surface conductivity. Insulating $AgNbO_3$ has considerable DOS only at around 2.8 eV. This agrees with the work of Lin *et al.*, where covalent materials with large valence bandwidths do not favor the final-state effect. [9] The exact number of electrons from $Nb^{4+}$ ($4d^1$) contribute to semiconducting $NbO_x$ and metallic $SrNbO_3$, but $NbO_x$ has more DOS in the valence band near Fermi. Likewise, $Nb^{5+}$ (intrinsic) with a $4d^0$ state offers no electrons to the DOS, but it still shows the DOS in the valence band near the Fermi level. This is due to various defects in the thin film lattice, which offer more DOS to NbO$_x$ than SrNbO$_3$, similar to the mid-gap defect states reported in oxygen-deficient metal oxides. [31, 32] These states make the Nb$_2$O$_5$ (intrinsic) thin film sample semi-conducting and the extrinsic Nb$_2$O$_5$ and powder materials insulating. The appearance of Nb$^{x+}$ in NbO$_x$ originates from these defect states near the Fermi level, which is also present for the (almost) fully oxidized sample Nb_0.35. Various surface and interfacial effects and defects can cause doping, making an insulator semiconducting when in thin film form. Even though SrNbO$_3$ is a better-conducting oxide than any NbO$_x$ samples in this work, we observed a larger contribution to the final-state effect from NbO$_x$ than that from SrNbO$_3$. This is due to the availability of more DOS in NbO$_x$ compared to SrNbO$_3$ for taking part in the core hole-valence electron coupling.

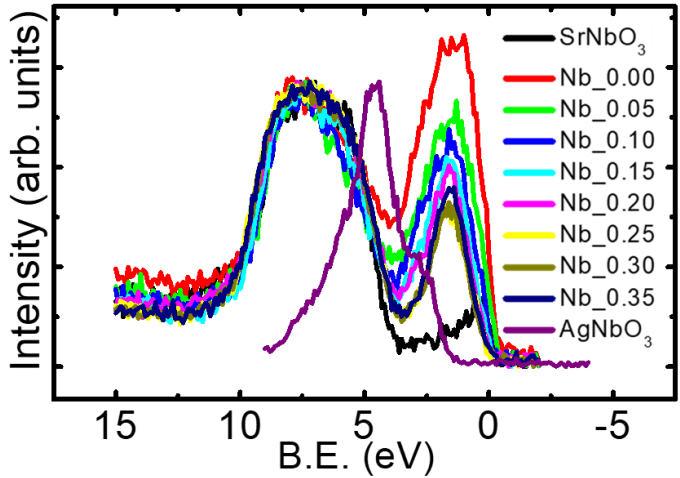

**Figure 2.** Surface valence band XPS spectra of various Nb-based samples

Keeping these understandings of final-state effects in mind, we fitted the XPS curves using the software CasaXPS. [33] Spectral lines of Nb$^0$, Nb$^{4+}$, Nb$^{5+}$ together with final-state spectral lines originating from both Nb$^{4+}$ and Nb$^{5+}$ were considered while deconvoluting the experimental data. A Shirley background was applied. We initiated the fit by giving a Δ value of 2.76 eV but later redefined it to 2.72 eV for better-fit results. The Nb 3d XPS spectra of NbO$_x$, together with fitted spectra, are shown in supplementary **Figure** S2, and the details are tabulated in **Table** S1. The full-width half maximum (FWHM) of spectral lines from the final-state effect is less than the FWHM of spectral lines of core-level, as in the case of plasmon satellites. [21]

**Figures** S3 (a) and (b) give the evolution of the concentration of $Nb^{5+}$ (initial state) and $Nb^{5x+}$ (final state), respectively, as a function of oxygen flow during growth. All the samples possess $Nb^{5+}$, either extrinsic or intrinsic, or both. We found metallic $Nb^0$ in samples Nb_0.00 to Nb_0.10 and believe that these samples possess respective amounts of extrinsic $Nb^{5+}$ contribution. The red line in **Figure** S3 (a) indicates an approximate oxygen flow level where all the metallic $Nb^0$ got converted into its oxides, $Nb^{4+}$ and/or $Nb^{5+}$. This line is also indicative of the absence of extrinsic $Nb^{5+}$, corresponding to the absence of metallic Nb. On the left side of this red line, the oxidations $Nb^0 \rightarrow Nb^{4+}$ and $Nb^0/Nb^{4+} \rightarrow Nb^{5+}$ happen. The black arrow in this region is a guide to the eye, indicating the decrease of extrinsic $Nb^{5+}$. On the right-hand side of the red line, only the oxidation $Nb^{4+} \rightarrow Nb^{5+}$ takes place, as there is no more metallic $Nb^0$ left. The contribution of the final-state spectral line is only from intrinsic $Nb^{5+}$, and no such contribution is expected from extrinsic native oxide formed on the surface of metallic Nb. The intrinsic $Nb^{5+}$ increases throughout the sample upon increasing the flow of oxygen and is apparent in the evolution of final-state $Nb^{5x+}$, as shown in **Figure** S3 (b). **Figures** S4 (a) and (b) depict the evolution of the concentration of initial-state $Nb^{4+}$ and final-state $Nb^{4x+}$ with oxygen flow during the growth of $NbO_x$. The red line in **Figure** S4 (a) indicates the cessation of metallic $Nb^0$, correspondingly the absence of extrinsic $Nb^{5+}$ as in the case of **Figure** S3 (a). Only intrinsic $Nb^{4+}$ is present in the samples (no extrinsic contribution to $Nb^{4+}$). The black arrows represent the evolution of intrinsic $Nb^{4+}$. The $Nb^{4+}$ is declining on the right-hand side of the red line due to the further oxidation of $Nb^{4+}$ into $Nb^{5+}$. A corresponding trend is observed in the final-state effect associated with $Nb^{4+}$ (**Figure** S4 (b)).

## 4. Conclusions

In summary, we fabricated thin films of partially oxidized $NbO_x$ samples and compared their XPS spectra with that of metallic conducting $SrNbO_3$ and insulating $AgNbO_3$. By taking the $Nb^{5+}$ spectral line from $AgNbO_3$ and native $Nb_2O_5$ formed on metallic Nb, we fixed the position of the $Nb^{5+}$ spectral line. The spectral line of $Nb^{4+}$ is fixed from that of $SrNbO_3$. To tackle the extra peak $Nb^x$ found at the lower binding energy side of Nb 3d XPS spectra of $SrNbO_3$, we compared it with that of $NbO_x$ samples. We found that the $Nb^x$ line is shifting towards higher binding energy, corresponding to the shift in $Nb^{4+}$ towards higher binding energy (towards the line of $Nb^{5+}$) in $NbO_x$ samples. This was confirmative of the formation of final-state spectral lines corresponding to both $Nb^{4+}$ and $Nb^{5+}$ in $NbO_x$ samples. With the deconvolution of the spectral lines, we found that the development of final-state effects is strongly linked to the formation of intrinsic oxides. In other words, DOS near the Fermi level is mandatory for such an effect to appear. The final-state effect is strikingly different from $Nb^{4+}$ and $Nb^{5+}$ as it is directly related to the number of charge carriers taking part in the core hole-valence electron coupling. Only the initial-state effect is exhibited by an oxide containing $Nb^{5+}$ when the DOS is absent near the Fermi level. However, when a thin film lattice offers defect states in the valence band close to Fermi energy, the binding energy of the spectral line is decided not only by core-level energy but also by the interaction of these valence states with the core hole, as in the case of metallic $SrNbO_3$. This results in manifold peaks in the XPS spectra of Nb 3d at lower binding energy than that of the core level, as reported by Lin *et al.* [9] The final-state effect exhibits a one-to-one correspondence with the availability of DOS in the valence band near the Fermi level and has no direct relation with the bulk conductivity of the metallic and semiconducting thin films. Such final-state contributions are absent in insulators with no DOS in the valence band near the Fermi level. We believe this work will serve as a byword to researchers working in Nb-based and similar materials systems while interpreting their chemical states.

**Supporting Information**

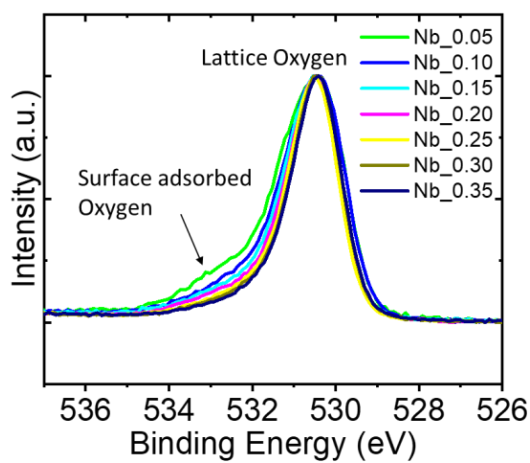

**Figure S1.** O 1s XPS spectra of $NbO_x$ samples prepared with different oxygen flows.

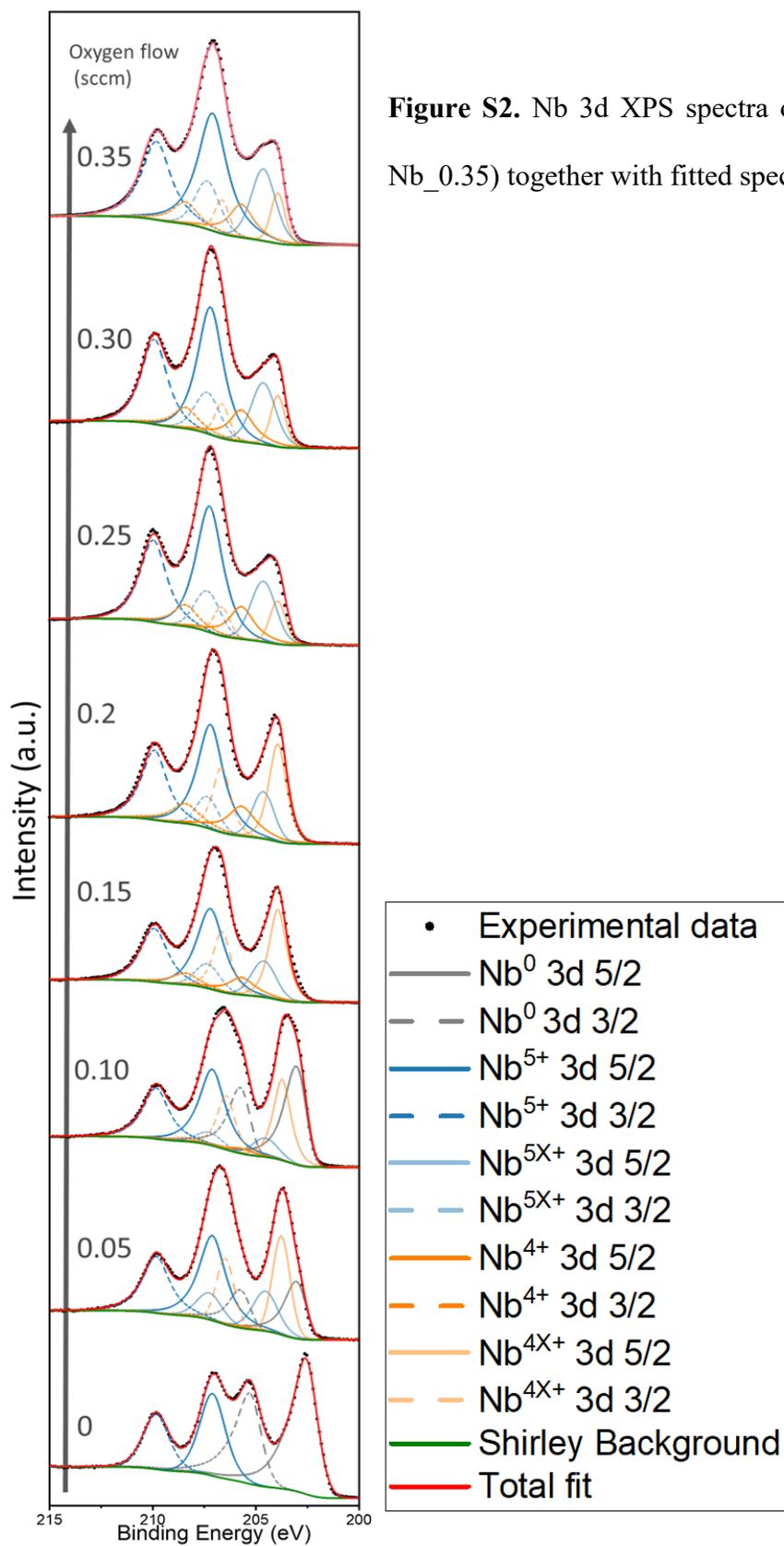

**Figure S2.** Nb 3d XPS spectra of NbO$_x$ (samples Nb_0.00 to Nb_0.35) together with fitted spectra.

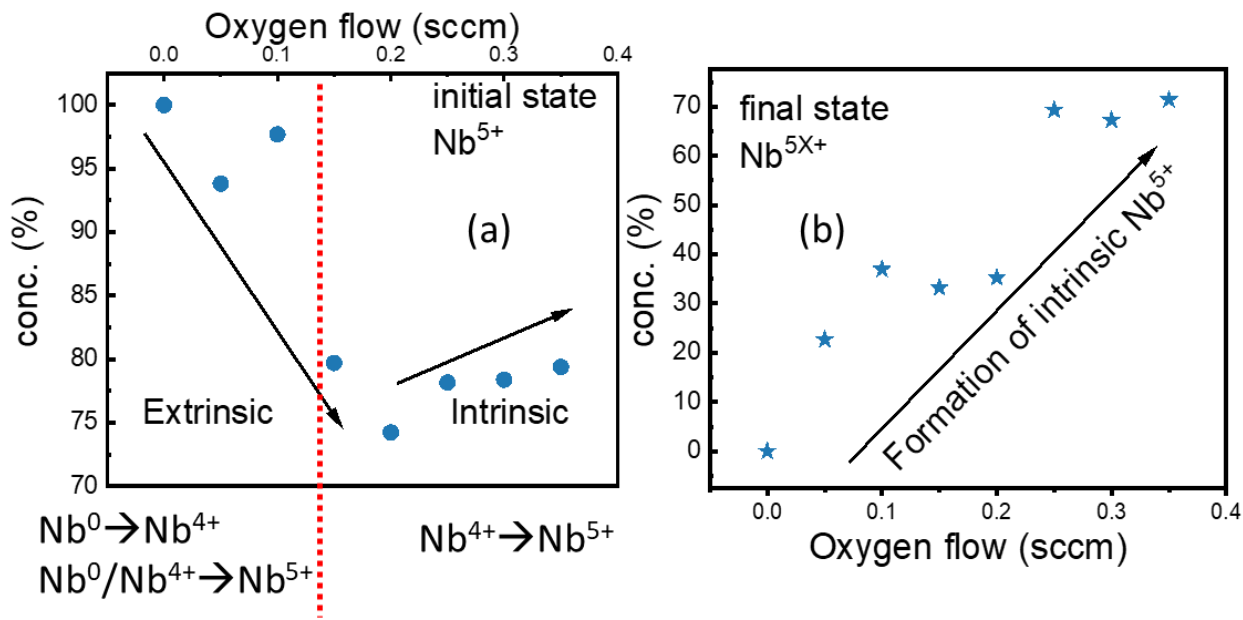

**Figure S3.** Evolution of the concentration of (a) initial-state $Nb^{5+}$ and (b) final-state $Nb^{5x+}$ with oxygen flow during the growth of $NbO_x$. The red line in (a) indicates the cessation of metallic $Nb^0$, correspondingly the absence of extrinsic $Nb^{5+}$. The black arrows in (a) represent the evolution of extrinsic and intrinsic initial-state $Nb^{5+}$. The black arrow in (b) represents the formation of intrinsic $Nb^{5+}$ since no final-state effect is associated with extrinsic $Nb^{5+}$.

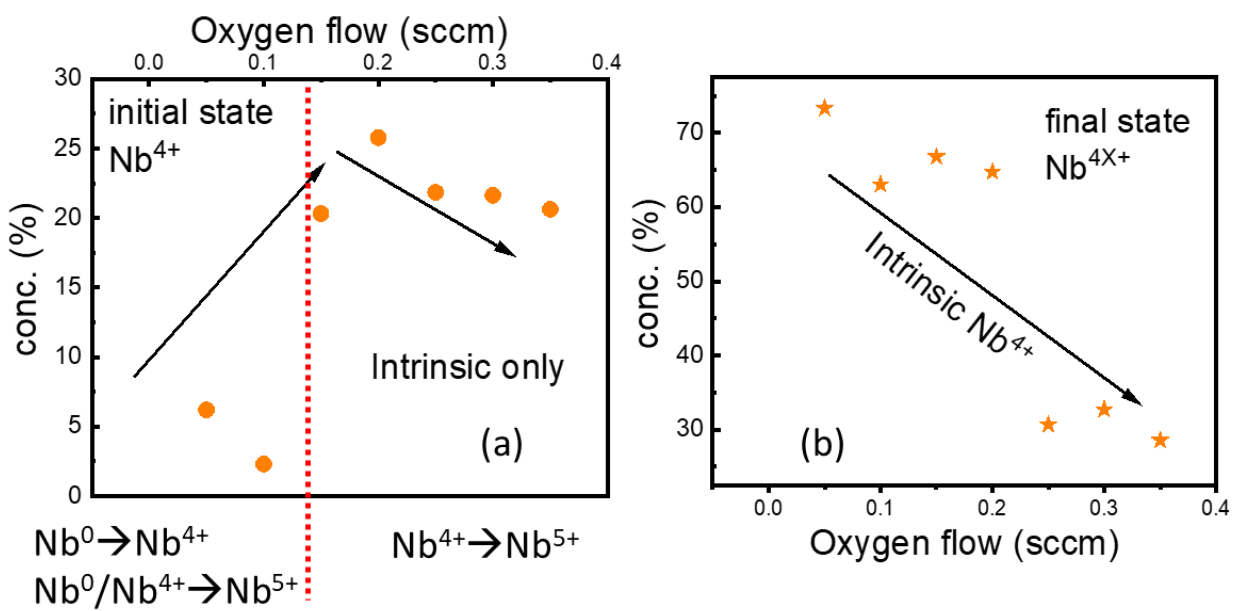

**Figure S4.** Evolution of the concentration of (a) initial-state $Nb^{4+}$ and (b) final-state $Nb^{4x+}$ with oxygen flow during the growth of $NbO_x$. The red line in (a) indicates the cessation of metallic $Nb^0$, correspondingly the absence of extrinsic $Nb^{5+}$. Only intrinsic $Nb^{4+}$ is present in the samples (no extrinsic contribution to $Nb^{4+}$). The back arrows in (a) and (b) represent the evolution of intrinsic $Nb^{4+}$.

**Table S1.** The details used/obtained during the deconvolution of Nb XPS spectra using CasaXPS.[33]

| | $Nb^0$ $3d_{5/2}$ | $Nb^{5+}$ $3d_{5/2}$ (Initial-state) | $Nb^{4+}$ $3d_{5/2}$ (Initial-state) | $Nb^{5x+}$ $3d_{5/2}$ (Final-state) | $Nb^{4x+}$ $3d_{5/2}$ (Final-state) |
|---|---|---|---|---|---|
| Line Shape | Gaussian/Lorentzian (GL) modified by an asymmetric form A (A GL) | Gaussian/Lorentzian (GL) | Gaussian-Lorentzian (GL) | Gaussian-Lorentzian (GL) | Gaussian-Lorentzian (GL) |
| Spin-orbit splitting ($\Delta$) (eV) | 2.72 | | | | |
| C 1s (eV) | 284.8 | | | | |
| Background | Shirley | | | | |
| **Sample Nb_0.00 (STD: 2.19)** | | | | | |
| B.E. (eV) | 202.58 | 207.10 | - | - | - |
| % | 58.28 | 41.72 | - | - | - |
| FWHM (eV) | 1.28 | 1.28 | - | - | - |
| Asymmetry index | | 0 | - | - | - |
| Fit Functions | A(0.32, 0.62, 6)(GL(30) | GL(92) | - | - | - |
| **Sample Nb_0.05 (STD: 2.89)** | | | | | |
| B.E. (eV) | 202.79 | 207.12 | 205.70 | 204.50 | 203.69 |
| % | 35.37 | 33.49 | 2.22 | 6.56 | 22,37 |
| FWHM (eV) | 1.085 | 1.47 | 1.43 | 0.99 | 1.08 |
| Asymmetry index | 99 | 0 | 0 | 0 | 0 |
| Fit function | LA(1.2,20,99) | GL(90) | GL(90) | GL(24) | GL(97) |

**Sample Nb_0.10 (STD: 2.38)**

| B.E. (eV) | 202.79 | 207.11 | 205.70 | 204.50 | 203.73 |
|---|---|---|---|---|---|
| % | 20.57 | 31.1 | 0.92 | 14.56 | 24,85 |
| FWHM (eV) | 1.09 | 1.55 | 1.43 | 1.43 | 1.08 |
| Asymmetry index | 99 | 0 | 0 | 0 | 0 |
| Fit function | LA(1.2, 20, 99) | GL(90) | GL(97) | GL(40) | GL(10) |

**Sample Nb_0.15 (STD: 4.77)**

| B.E. (eV) | - | 207.20 | 205.70 | 204.58 | 203.89 |
|---|---|---|---|---|---|
| % | - | 42.95 | 10.95 | 15.30 | 30.81 |
| FWHM (eV) | - | 1.53 | 1.54 | 1.40 | 1.01 |
| Asymmetry index | - | 0 | 0 | 0 | 0 |
| Fit function | - | GL(82) | GL(98) | GL(10) | GL(75) |

**Sample Nb_0.20 (STD: 4.79)**

| B.E. (eV) | - | 207.20 | 205.70 | 204.60 | 203.90 |
|---|---|---|---|---|---|
| % | - | 37.57 | 16.75 | 15.24 | 30.43 |
| FWHM (eV) | - | 1.64 | 1.64 | 1.23 | 1.01 |
| Asymmetry index | - | 0 | 0 | 0 | 0 |
| Fit function | - | GL(88) | GL(97) | GL(10) | GL(65) |

**Sample Nb_0.25 (STD: 4.90)**

| B.E. (eV) | - | 207.25 | 205.70 | 204.60 | 203.90 |
|---|---|---|---|---|---|
| % | - | 54.34 | 14.66 | 21.42 | 9.57 |
| FWHM (eV) | - | 1.52 | 1.44 | 204.6 | 0.89 |
| Asymmetry index | - | 0 | 0 | 0 | 0 |
| Fit function | - | GL(80) | GL(97) | GL(5) | GL(3) |

**Sample Nb_0.30 (STD: 3.00)**

| B.E. (eV) | - | 207.20 | 205.70 | 204.60 | 203.90 |
|---|---|---|---|---|---|
| % | - | 52.97 | 14.61 | 21.81 | 10.61 |
| FWHM (eV) | - | 1.47 | 1.44 | 1.40 | 0.81 |
| Asymmetry index | - | 0 | 0 | 0 | 0 |
| Fit function | - | GL(82) | GL(97) | GL(40) | GL(40) |

**Sample Nb_0.35 (STD: 2.63)**

| B.E. (eV) | - | 207.20 | 205.70 | 204.60 | 203.90 |
| --- | --- | --- | --- | --- | --- |
| % | - | 53.66 | 13.94 | 23.15 | 9.26 |
| FWHM (eV) | - | 1.47 | 1.44 | 1.40 | 0.81 |
| Asymmetry index | - | 0 | 0 | 0 | 0 |
| Fit Function | - | GL(92) | GL(97) | GL(40) | GL(20) |

**Acknowledgement**

This work was supported by Deutsche Forschungsgemeinschaft (DFG) under project 429646908. JP and PH are thankful for the funding received from DFG Collaborative Research Center (CRC) 1073 and Ministerium für Wissenschaft, Forschung und Kunst (MWK), Lower Saxony 'Zukunft. Niedersachsen'. JP and LA acknowledge the financial support from CRC/TRR 270 and CRC 1487. The authors thank Mr. Thorsten Schneider, TU Darmstadt, for the XPS spectra of $AgNbO_3$. JP thanks Ms. Gabriele Haindl, TU Darmstadt, for the technical help received during the thin film deposition.

**Author Contributions**

J.P. - Designed and performed the experiments. Collected and analyzed the data. Wrote and edited the manuscript. Funding acquisition.

P.H. – Analyzed the data and performed the curve fitting. Wrote and edited the manuscript.

L.A. - Analyzed the data. Wrote and edited the manuscript. Funding acquisition.

**References**

[1] Y. Lian, N. Yang, D. Wang, Y. Zheng, C. Ban, J. Zhao, H. Zhang, Advanced Energy and Sustainability Research, 1 (2020) 2000038.
[2] R. Muccillo, E.N.S. Muccillo, International Journal of Ceramic Engineering & Science, 6 (2024) e10201.
[3] C. Laverick, Journal of the Less Common Metals, 139 (1988) 107-122.
[4] M.A. Melo, I.M. Brito, J.V.S.B. Mello, P.S.M. Rocha, I.A.A. Bessa, B.S. Archanjo, F.S. Miranda, R.J. Cassella, C.M. Ronconi, ChemCatChem, n/a (2023) e202300387.
[5] C. Liu, B. Wang, L. Xu, K. Zou, W. Deng, H. Hou, G. Zou, X. Ji, ACS Applied Materials & Interfaces, 15 (2023) 5387-5398.
[6] S. Zhang, J. Hwang, Y. Sato, K. Matsumoto, R. Hagiwara, ACS Applied Energy Materials, 6 (2023) 2333-2339.
[7] E.C. Wheeler-Jones, M.J. Loveridge, R.I. Walton, Batteries & Supercaps, 6 (2023) e202200556.
[8] P.v.d. Heide, X-Ray Photoelectron Spectroscopy: An Introduction to Principles and Practices, John Wiley & Sons, 2011.
[9] C. Lin, A. Posadas, T. Hadamek, A.A. Demkov, Physical Review B, 92 (2015) 035110.

[10] D.O. Scanlon, G.W. Watson, D.J. Payne, G.R. Atkinson, R.G. Egdell, D.S.L. Law, The Journal of Physical Chemistry C, 114 (2010) 4636-4645.
[11] R. Fontaine, R. Caillat, L. Feve, M.J. Guittet, Journal of Electron Spectroscopy and Related Phenomena, 10 (1977) 349-357.
[12] A.B. Posadas, A. O'Hara, S. Rangan, R.A. Bartynski, A.A. Demkov, Applied Physics Letters, 104 (2014) 092901.
[13] A.R. Dhamdhere, T. Hadamek, A.B. Posadas, A.A. Demkov, D.J. Smith, Journal of Applied Physics, 120 (2016) 245302.
[14] J.K. Hulm, C.K. Jones, R.A. Hein, J.W. Gibson, Journal of Low Temperature Physics, 7 (1972) 291-307.
[15] T.B. Reed, E.R. Pollard, L.E. Lonney, R.E. Loehman, J.M. Honig, Niobium Monoxide, in: Inorganic Syntheses, 1995, pp. 108-110.
[16] J. Halbritter, Applied Physics A, 43 (1987) 1-28.
[17] S. Thapa, S.R. Provence, P.T. Gemperline, B.E. Matthews, S.R. Spurgeon, S.L. Battles, S.M. Heald, M.A. Kuroda, R.B. Comes, APL Materials, 10 (2022) 091112.
[18] T. Ofoegbuna, P. Darapaneni, S. Sahu, C. Plaisance, J.A. Dorman, Nanoscale, 11 (2019) 14303-14311.
[19] C. Bigi, P. Orgiani, J. Sławińska, J. Fujii, J.T. Irvine, S. Picozzi, G. Panaccione, I. Vobornik, G. Rossi, D. Payne, F. Borgatti, Physical Review Materials, 4 (2020) 025006.
[20] H. Wadati, J. Mravlje, K. Yoshimatsu, H. Kumigashira, M. Oshima, T. Sugiyama, E. Ikenaga, A. Fujimori, A. Georges, A. Radetinac, K.S. Takahashi, M. Kawasaki, Y. Tokura, Physical Review B, 90 (2014) 205131.
[21] P.A.D. Gonçalves, F.J. García de Abajo, Nanoscale, 15 (2023) 11852-11859.
[22] M. Mirjolet, M. Kataja, T.K. Hakala, P. Komissinskiy, L. Alff, G. Herranz, J. Fontcuberta, Advanced Optical Materials, n/a (2021) 2100520.
[23] F.A. Stevie, R. Garcia, J. Shallenberger, J.G. Newman, C.L. Donley, Journal of Vacuum Science & Technology A, 38 (2020) 063202.
[24] M. Ishfaq, M. Rizwan Khan, M.F. Bhopal, F. Nasim, A. Ali, A.S. Bhatti, I. Ahmed, S. Bhardwaj, C. Cepek, Journal of Applied Physics, 115 (2014) 174506.
[25] B. Yang, J. Bian, L. Wang, J. Wang, Y. Du, Z. Wang, C. Wu, Y. Yang, Physical Chemistry Chemical Physics, 21 (2019) 11697-11704.
[26] T.J. Frankcombe, Y. Liu, Chemistry of Materials, 35 (2023) 5468-5474.
[27] A.K.-V. Alexander V. Naumkin, Stephen W. Gaarenstroom, and Cedric J. Powell, in.
[28] X.-r.P.S.X.R. Pages, in.
[29] D.V. Louzguine-Luzgin, T. Hitosugi, N. Chen, S.V. Ketov, A. Shluger, V.Y. Zadorozhnyy, A. Caron, S. Gonzales, C.L. Qin, A. Inoue, Thin Solid Films, 531 (2013) 471-475.
[30] A.S. Trifonov, A.V. Lubenchenko, V.I. Polkin, A.B. Pavolotsky, S.V. Ketov, D.V. Louzguine-Luzgin, Journal of Applied Physics, 117 (2015) 125704.
[31] N. Kaiser, T. Vogel, A. Zintler, S. Petzold, A. Arzumanov, E. Piros, R. Eilhardt, L. Molina-Luna, L. Alff, ACS Applied Materials & Interfaces, 14 (2022) 1290-1303.
[32] X. Chen, L. Liu, P.Y. Yu, S.S. Mao, Science, 331 (2011) 746-750.
[33] N. Fairley, V. Fernandez, M. Richard-Plouet, C. Guillot-Deudon, J. Walton, E. Smith, D. Flahaut, M. Greiner, M. Biesinger, S. Tougaard, D. Morgan, J. Baltrusaitis, Applied Surface Science Advances, 5 (2021) 100112.